\documentclass{elsart}
\usepackage{epsfig}
\begin{document}
\title{Information theory approach (extensive and nonextensive) to
high energy multiparticle production processes} 
\author{F.S.Navarra$^{a}$, O.V.Utyuzh$^{b}$, G. Wilk$^{b}$, 
Z.W\l odarczyk$^{c}$} 
\address{$^{a}${\it Instituto de F\'{\i}sica, Universidade de S\~{a}o
               Paulo\\
               CP 66318, 05389-970 S\~{a}o Paulo, SP, Brazil; e-mail:
               navarra@if.usp.br}\\
         $^{b}$The Andrzej So\l tan Institute for Nuclear Studies\\
                Ho\.za 69; 00-689 Warsaw, Poland; e-mail: wilk@fuw.edu.pl\\
         $^{c}$Institute of Physics, \'Swi\c{e}tokrzyska Academy,\\
               \'Swi\c{e}tokrzyska 15; 25-406 Kielce, Poland; e-mail:
               wlod@pu.kielce.pl\\ 
 \today}
{\scriptsize Abstract: We present an overview of information theory
approach (both in its extensive and nonextensive versions) applied to
high energy multiparticle production processes. It will be
illustrated by analysis of single particle distributions measured in
proton-proton, proton-antiproton and nuclear collisions. We shall
demonstrate the particular role played by the nonextensivity
parameter $q$ in such analysis as summarizing our knowledge on the
fluctuations existed in hadronizing system. 


}
\section{Introduction}

Information theory has been established to be extremely useful tool
in many branches of science \cite{INFO}. Its first application to
multiparticle production processes is very old \cite{Chao}, however,
it was used only sporadically since then \cite{MaxEnt,qMaxEnt}. Here
we shall present its methods and analyse, as example, single particle
spectra obtained in multiparticle production processes at high energy
\cite{MaxEnt,qMaxEnt}. Such processes are responsible for the
majority  of observed cross section. If $k_{1,2}$ are four-momenta of
colliding objects and $W=\sqrt{s} = \sqrt{\left(k_1 +
k_2\right)^2}$ is  invariant energy of collision, then the mean
multiplicity of produced secondaries increases with energy like
$\langle N\rangle \propto  \ln^2 W$ or $\langle N\rangle \propto
W^{\alpha}$ (with $\alpha \sim 0.5$) reaching in heavy ion collisions
values of the order of $\sim 10^3 \div \sim 10^4$. This fact makes
analysis of such processes very difficult, in particular it is
impossible to describe them from {\it first principles}. On the other
hand, understanding these processes is very important because they
provide background for more sophisticated measurements and they are
themselves valuable source of information on the hadronization
process in which initial energy $W$ converts into observed
secondaries (mostly mesons $\pi$ and $K$). One is looking therefore
for phenomenological models and first candidates were statistical
models of different kinds \cite{F,HAG}. With necessary modifications
and improvements they are actually in common use also nowadays and are
very successful \cite{NEWSTAT}. They all assume some kind of local
thermal equilibrium forming at early stage of collision process.
However, this assumption is not necessary because the characteristic
"thermal-like" (i.e., exponential) behaviour of some observables
arises always whenether one restricts itself to observation of only a
small part of some large system \cite{RES}. This is precisely
situation encountered in multiparticle production processes where
usually only a small part of produced secondaries is registered by
detectors and out of them only very limited number is subjected to
final analysis. The necessary averaging over all unmeasured degrees
of freedom introduces therefore a kind of {\it randomization} or a
{\it heat bath}, action of which can be summarily described by a
single parameter $T$, a kind of "effective temperature" of the usual
thermodynamics \cite{RES}. It could, however, happen that such
"thermal bath" is more complicated (exhibiting, for example, some
intrinsic fluctuations or collective flows) \cite{Q}. In such cases  
parameter $T$ must be supplemented by additional parameter summarily
describing the action of these new factors - in this way the
nonextensivity parameter $q$ appears (such "heat baths" are not
extensive anymore) \cite{Q} and notion of nonextensive statistics
enters in a natural way \cite{T}.  

Why use information theory methods in analysing multiparticle production 
processes? To illustrate this let us suppose that experiment measures 
some new distribution of secondaries. Immediately theoreticians rush
to describe the new data and provide a number of models, which differ
drastically in what concerns their assumptions, nevertheless they all
do explain experimental results in a remarkable good way. What has
happened? To answer this one has to introduce notion of {\it
information}: data considered contain only limited amount of
information and all models possess it as well. They differ therefore
in what concerns additional redundant information they posses, which
reflects not so much the state of the measured object but rather the
state of minds of scientists proposing these models. To quantify this
one has to resort to methods of information theory and introduce
information entropy as a measure of information \cite{INFO,Chao,MaxEnt}.   
The problem one is facing can be formulated in following way.
Suppose one has system {\bf S}, depending on some variable $x\in
(x_{min},x_{max})$, on which one performs finite number $n$
measurements providing us with $R_{1,\dots,n}$ results. The question
is: how to obtain {\it the most plausible} ({\it least biased}) and
model independent probability distribution, $p(x)$, of variable $x$?
The answer is \cite{INFO,MaxEnt}: look for such $p(x)$ (normalized to
unity) which maximizes information entropy under constraints given by
results of measurements. To account for many known features of the
system under consideration (like long-range correlations,
fractal-like structure of the phase space or intrinsic fluctuations,
which will be of special interest to us later \cite{WW}) we shall use
Tsallis entropy \cite{T} characterised by parameter $q$, which for $q
\rightarrow 1$ becomes the usual Shannon entropy \cite{INFO} 
\begin{equation}
S_q\, =\, - \frac{1}{1-q}\, \left[ 1\, -\, \int_{x_{min}}^{x_{max}}\,
dx\, p(x)^q\right]\quad \stackrel{q\rightarrow 1}{\Longrightarrow}\quad
- \int_{x_{min}}^{x_{max}}\,dx\, p(x)\, \ln p(x) , \label{eq:infoent}
\end{equation}
under constraints\footnote{For our limited purposes this form is
adequate. Formalism using escort probability distributions, $P_i =
p_i^q/\sum_i p_i^q$, leads to distributions of the type $c\left[1 -
(1-q)x/l\right]^{q/(1-q)}$, which is {\it formally identical} with
(\ref{eq:formulaq}), $c\left[1 - (1-Q)x/L\right]^{1/(1-Q)}$, provided
we identify: $Q = 1 + (q-1)/q$,  $L=l/q$ and $c=(2-Q)/L=1/l$. Now
$\langle x\rangle = L/(3-2Q) = l/(2-q)$ and $0<Q<1.5$ (to be compared
with $0.5< q < 2$).  Both distributions are identical and the
problem, which of them better describes data is at this level of
sophistication not important.}
\begin{equation}
\int_{x_{min}}^{x_{max}}\, dx\, R^{(k)}(x)\cdot
\left[p(x)\right]^q\, = \langle R^{(k)}\rangle_q =
R^{(q)}_{k=1,\dots,n} . \label{eq:constr} 
\end{equation}
As result one gets
\begin{equation}
p(x)\, =\, \frac{1}{Z_q} \exp_q\left[ -\, \sum_{k=1}^n  
        \beta^{(q)}_k\cdot R^{(k)}(x)\right] ,\label{eq:formulaq}
\end{equation}
where  
\begin{equation}
\exp_q\left( \frac{x}{\Lambda}\right)\, =\, \left[1 +
                   (1-q)\frac{x}{\Lambda}\right]^{1/(1-q)}\qquad
                   \stackrel{q \rightarrow 1}{\Longrightarrow}\qquad
                    \exp (\frac{x}{\Lambda})  . \label{eq:def} 
\end{equation}
Here $\beta^{(q)}_k$ are Lagrange multipliers obtained by solving
constraint equations and $Z_q$ ensures normalization of $p(x)$ to 
unity. It should be mentioned here that fluctuations in the parameter
$1/\Lambda$ of the exponential distribution lead to the
$q$-exponential with $q$ given by normalized variation of the
parameter $1/\Lambda$ \cite{WW,BeckC}. 

Long time ago it was shown using this method (with $q=1$)
\cite{Chao} that in order to fit single particle distributions
$dN/dy$ in rapidity variable $y=\frac{1}{2}\ln(E + p_L)/\ln(E -
p_L)$\footnote{~Here $E$ is energy of the observed particle produced
in collision process and $p_L$ is its longitudinal momentum, i.e.,
projection of its total momentum on the collision axis. The respective
transverse momenta $p_T$ are integrated over and enter only via their
mean value, $\langle p_T\rangle$, defining the so called transverse
mass of the particle: $\mu_T = \sqrt{\mu^2 + \langle p_T\rangle^2}$
with $\mu$ denoting its real mass. Notice that $E=\mu_T \cosh y$
whereas $p_L = \mu_T \sinh y$.} and multiparticle distributions
$P(N;\bar{N})$ measured at that time it was enough to assume that:
$(i)$  transverse momenta of produced secondaries $p_T$ are
limited  (i.e., phase space is effectively one dimensional) and
$(ii)$ not the whole available energy $\sqrt{s}$ is used for
production of the observed secondaries but only its fraction $K\in
(0,1)$ called inelasticity, the rest of energy is taken away by the
so called "leading particles". Closer scrutiny of numerous models
competing at that time revealed that {\it all of them} contained
these two assumptions (explicitly or implicitly) and that these
assumptions consisted the only part {\it common to all of them}. All
other additional assumptions were spurious and unnecessary and as
such could be safely abandoned without spoiling agreement with data. 

\section{Results}

\begin{figure}[ht]
  \begin{minipage}[ht]{54mm}
    \centerline{
        \epsfig{file=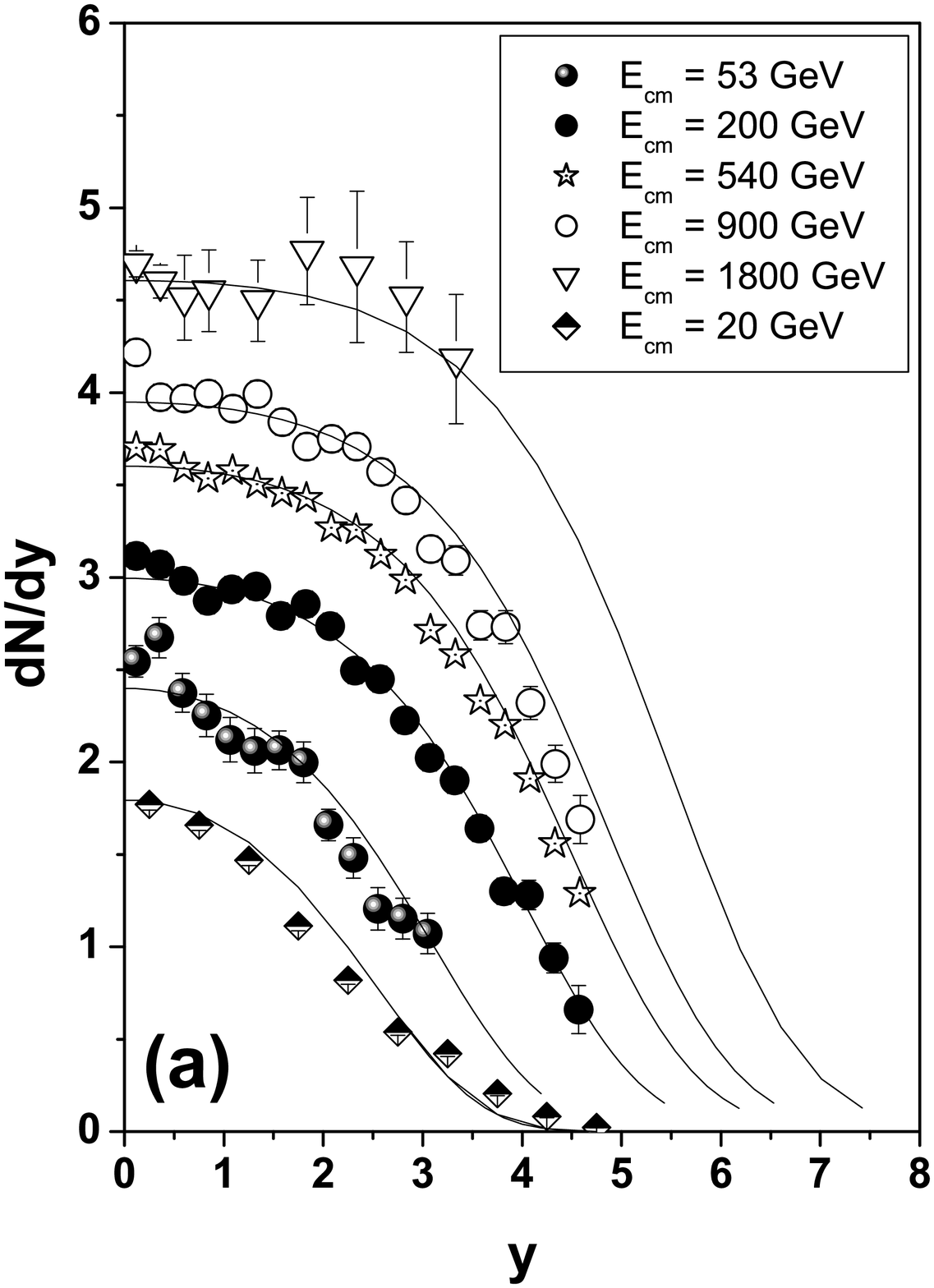, width=53mm}
     }
  \end{minipage}
\hfill
  \begin{minipage}[ht]{54mm}
    \centerline{
        \epsfig{file=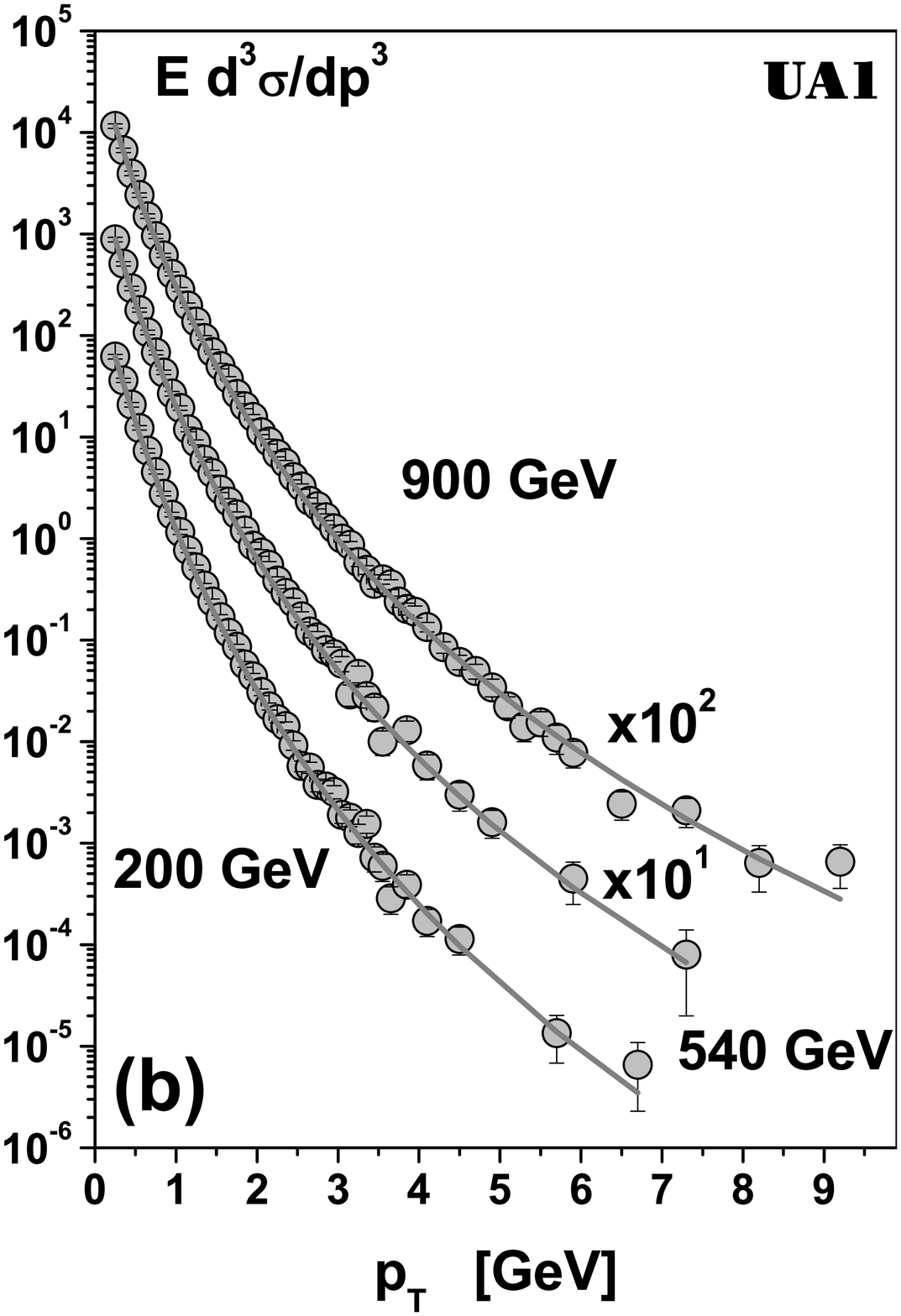, width=50mm}
     }
  \end{minipage}
\hfill
  \begin{minipage}[ht]{55mm}
    \centerline{
       \epsfig{file=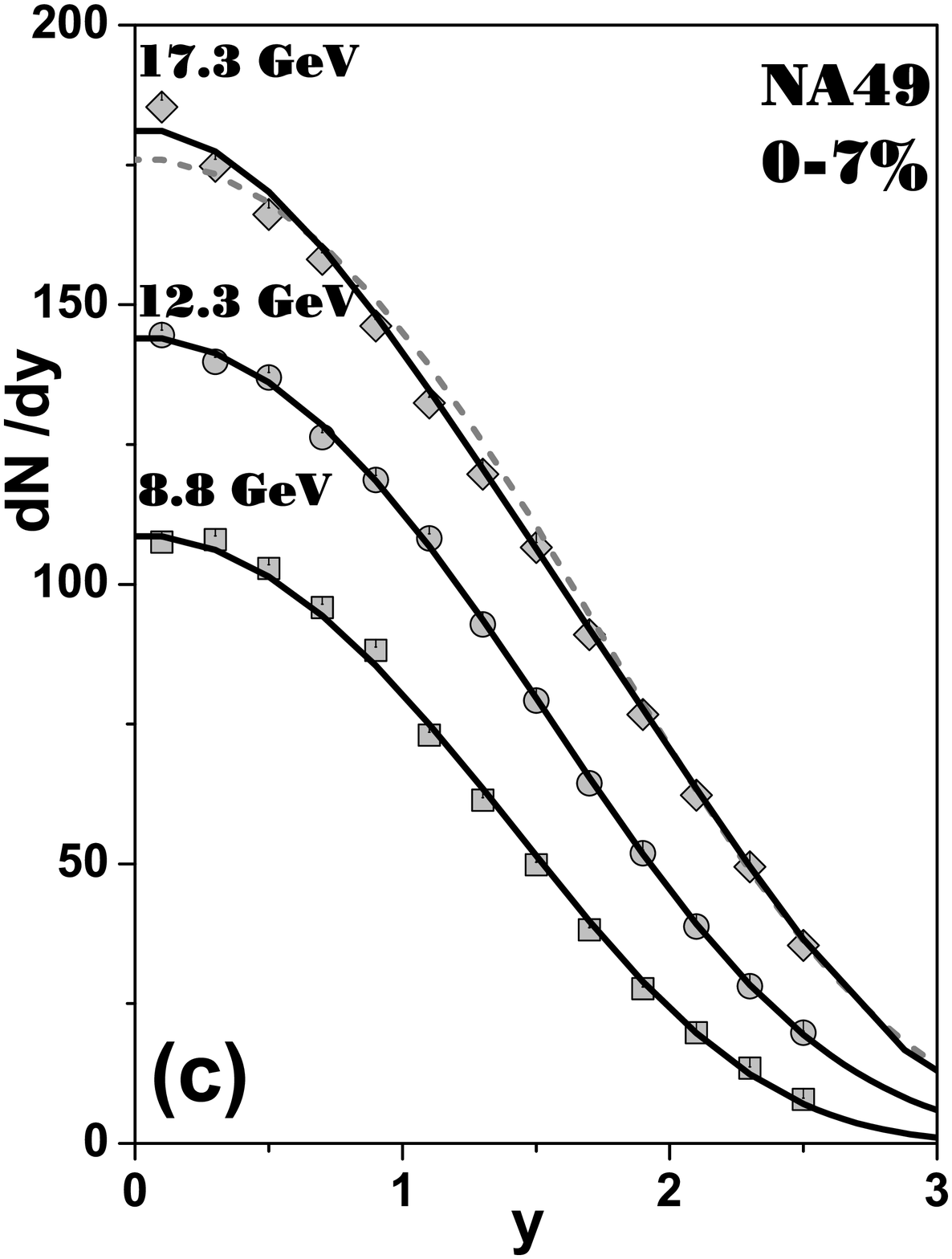, width=55mm}
     }
  \end{minipage}
  \caption{
\footnotesize Examples of applying eq. (\ref{eq:formulaq}). $(a)$ fit
to rapidity spectra for charged pions produced in $pp$ and 
$\bar{p}p$ collisions at different energies \cite{Datapp1}. $(b)$ fit
to $p_T$ spectra from UA1 experiment \cite{Datapt}. $(c)$ fits to
rapidity  spectra for negatively charged pions produced in $PbPb$
collisions at different energies (for the most central $0-7\%$
interval of impact parameter of collision \cite{DataAA}). The best
fit for $17.3$ GeV is for two-source case, cf. Fig.
\ref{fig:Figure2}f, with $q=1$ and $\Delta y = 0.83$. Total
inelasticity in this case is $K=0.58$. See text for other details.}      
  \label{fig:Figure1}
\end{figure}

We shall present now examples of analysis of single particle
distributions in rapidity for $pp$ and $p\bar{p}$ collisions at
energies varying between $\sqrt{s}=20$ GeV to $1800$ GeV
\cite{qMaxEnt} and first results for central nuclear collisions (see
Fig. \ref{fig:Figure1}). Our input consists of: $(i)$ the energy
of collision, $E_{cm}=\sqrt{s}$, out of which, according to result of
\cite{Chao}, we shall retain only fraction $K_q$ for production of
secondaries; $(ii)$ the mean multiplicity of charged secondaries
produced in nonsingle diffractive reactions at given energy:
$\bar{n}_{ch} = -7.0 + 7.2 s^{0.127}$ (out of which we construct the
total number of produced particles, $N=\frac{3}{2}\bar{n}_{ch}$
needed in our procedure); $(ii)$ the fact that phase space is
essentially one dimensional with mean transverse momentum slightly
energy dependent given by $\langle p_T\rangle = 0.3 +
0.044\ln\left(\sqrt{s}/20\right)$ \cite{data} (all secondaries will
be assumed to be pions of mass $\mu = 0.14$ GeV) and $(iv)$ that
multiplicity  distributions of produced secondaries is not Poissonian
(like it was in \cite{Chao}) but of Negative Binomial (NB) type
\begin{equation}
P(n)\, =\, P(n;\bar{n},k)\, =\, \frac{\Gamma(k+n)}{\Gamma(1+n)\Gamma(k)}\cdot
           \frac{\gamma^k}{(\gamma+1)^{k+n}}\qquad {\rm with}\qquad
\gamma =\frac{k}{\bar{n}} , \label{eq:NB}
\end{equation}
where parameter $k$ ($k \ge 1$) affects width of $P(n)$ and is given by
its normalized variance \cite{data}
\begin{equation}
\frac{1}{k}\, =\, \frac{\sigma^2\left(n_{ch}\right)}{\langle
n_{ch}\rangle^2} - \frac{1}{\langle n_{ch}\rangle}\, =\, -0.104 + 0.058
\ln \sqrt{s} .\label{eq:NBfit}
\end{equation}

This last point is new and most important because it can be accounted 
for only in nonextensive version of our method with $q>1$. 
This is because the value of $k^{-1}$ may be understood as the
measure of fluctuations of mean multiplicity caused by some dynamical
factors. When there are only statistical fluctuations in hadronizing
system then $P(n)$ is of Poissonian type. Intrinsic (dynamical)
fluctuations would mean fluctuating mean multiplicity $\bar{n}$. In
the case when such fluctuations are given by a gamma 
distribution with normalized variance $D(\bar{n})$ then, as a
result, one obtains the NB multiplicity distribution with $1/k =
D(\bar{n})$ (because convolution of Poissonian and Gamma
distributions results in NB one \cite{Shih}). Assuming now that these
fluctuations contribute to nonextensivity, i.e., that they define
parameter $q=1+D(\bar{n})$, one is lead to the use of $q>1$ (i.e.,
nonextensive formalism) with $q=1+1/k$, where $k$ is parameter
defining NB form of $P(n)$, energy dependence of which is given by.
eq. (\ref{eq:NBfit}). 

With such input the most probable rapidity distribution is 
\begin{equation}
p(y)\, =\, \frac{1}{N}\frac{dN}{dy}\, =\, \frac{1}{Z_q} \exp_q 
       \left( - \beta_q \cdot \mu_T \cosh y \right), \label{eq:py}
\end{equation}
where $Z_q$ and $\beta_q$ are not free parameters, like in similar
formulas used in applications of statistical models \cite{Chou}, but 
are given by the normalization condition and by the energy
conservation constraint ($\mu_T = \sqrt{\mu^2 + \langle
p_T\rangle^2}$, see \cite{qMaxEnt} for details):  
\begin{equation}
\int_{-Y_m}^{Y_m} \, dy\, \mu_T\cdot \cosh y \cdot [p(y)]^q\, =\,
\frac{\kappa_q \cdot \sqrt{s}}{N} .\label{eq:energy}
\end{equation}
Symmetry of the collision (identical particles) implies that momentum
is conserved in this case automatically. There are therefore two
parameters, nonextensivity $q$ responsible for fluctuations (as
mentioned above) and $q$-inelasticity $\kappa_q$ to be deduced
directly from data essentially in a {\it model independent 
way}, cf. Fig. \ref{fig:Figure1}a. There is, however,
question concerning physical meaning of $\kappa_q$ for $q>1$ case
considered here (notice that our $q$ changes from $1.05$ to $1.33$
when going from $\sqrt{s} = 20$ GeV to $1800$ GeV whereas the
"partition temperature" $T_q = 1/\beta_q$ changes, accordingly, from
$1.76$ GeV to $62.57$ GeV \cite{qMaxEnt}). The point is, as was
stressed in \cite{Beck}, that in case of $q>1$ our energy $\kappa_q$
contains also interaction between particles and is not directly seen
in rapidity distribution. However, the quantity (inelasticity) 
\begin{equation}
K_q = \frac{N}{\sqrt{s}}\langle E_q\rangle\, =\, \frac{N}{\sqrt{s}}
\cdot \int_{-Y_m}^{Y_m}dy\, p(y) \left[ \mu_T\cdot \cosh y \right]\,
\sim\, \frac{\kappa_q}{3-2q} \label{eq:kq}
\end{equation}
is the correct fraction of energy used for production of secondaries
particles (it happens to be fairly constant with $\sqrt{s}$ varying 
slightly around $K_q \sim 0.5$ \cite{qMaxEnt}).
Fig. \ref{fig:Figure1}b shows example of similar fit but this time to
the observed $p_T$ spectra at energies $200$, $500$ and $900$ GeV
\cite{Datapt} (using eq. (\ref{eq:py}) with energy $\mu_T \cosh y$
replaced by transverse momentum $p_T$). As can be noticed fits are
very good in the whole range of $p_T$ with the corresponding values
of $T_T=1/\beta_T$ (in GeV) and $q_T$ equal to $(T_T;q_T)= (0.134;1.095)$,
$(0.135;1.105)$ and $(0.14;1.11)$ for energies $200$, $500$ and
$900$, respectively. These values must be compared with the
corresponding values of $(T=T_L;q=q_L)$ for rapidity distributions,
which are equal to, respectively: $(11.74;1.2)$, $(20.39;1.26)$ and
$(30.79;1.29)$ at comparable energies  \cite{qMaxEnt}. 
The question then arises, what is the actual value of the parameter
$q$ describing fluctuations of the mean multiplicity, i.e.,
affecting, according to the previous discussion, multiplicity
distributions? It is sensitive to $p=\sqrt{p^2_L + p^2_T}$ and, as we
have seen here, both $p_L$ and $p_T$ show traces of fluctuations by
leading to $q>1$. At present we can offer only some approximate
answer because there are no data measuring $p_T$ distributions at all
values of rapidity $y$, i.e., providing correlations between parameters
$(T;q) = (T_L;q_L)$ for longitudinal momenta (rapidity) distributions
and  $(T;q) = (T_T;q_T)$ for transverse momenta distributions.
Noticing that $q-1 = \sigma^2(T)/T^2$ (i.e., it is given by
fluctuations of total temperature $T$) and assuming that $\sigma^2(T) =
\sigma^2(T_L) + \sigma^2(T_T)$, one can estimate that resulting values of
$q$ should not be too different from
\begin{equation}
q\, =\, \frac{q_L\, T_L^2\, +\, q_T\, T^2_T}{T^2}\, -\, 
        \frac{T^2_L\, +\, T^2_T}{T^2}\, +\, 1, 
\label{eq:qqq} 
\end{equation}
which, for $T_L \gg T_T$, as is in our case, leads to the result that
$q \sim q_L$, i.e., it is given by the longitudinal (rapidity)
distributions only. 

\begin{figure}[ht]
  \begin{minipage}[ht]{55mm}
    \centerline{
        \epsfig{file=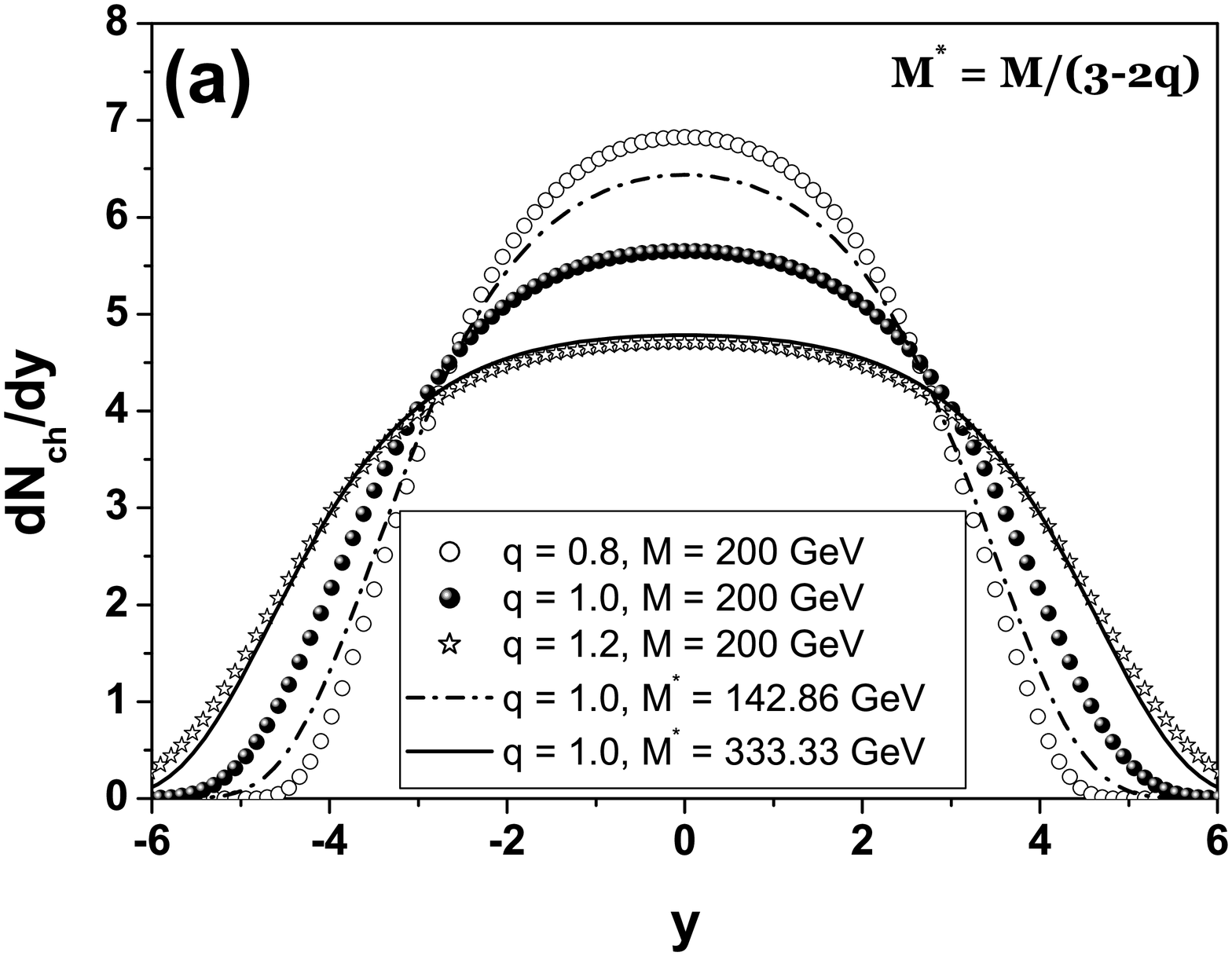, width=55mm}
     }
  \end{minipage}
\hfill
  \begin{minipage}[ht]{55mm}
    \centerline{
        \epsfig{file=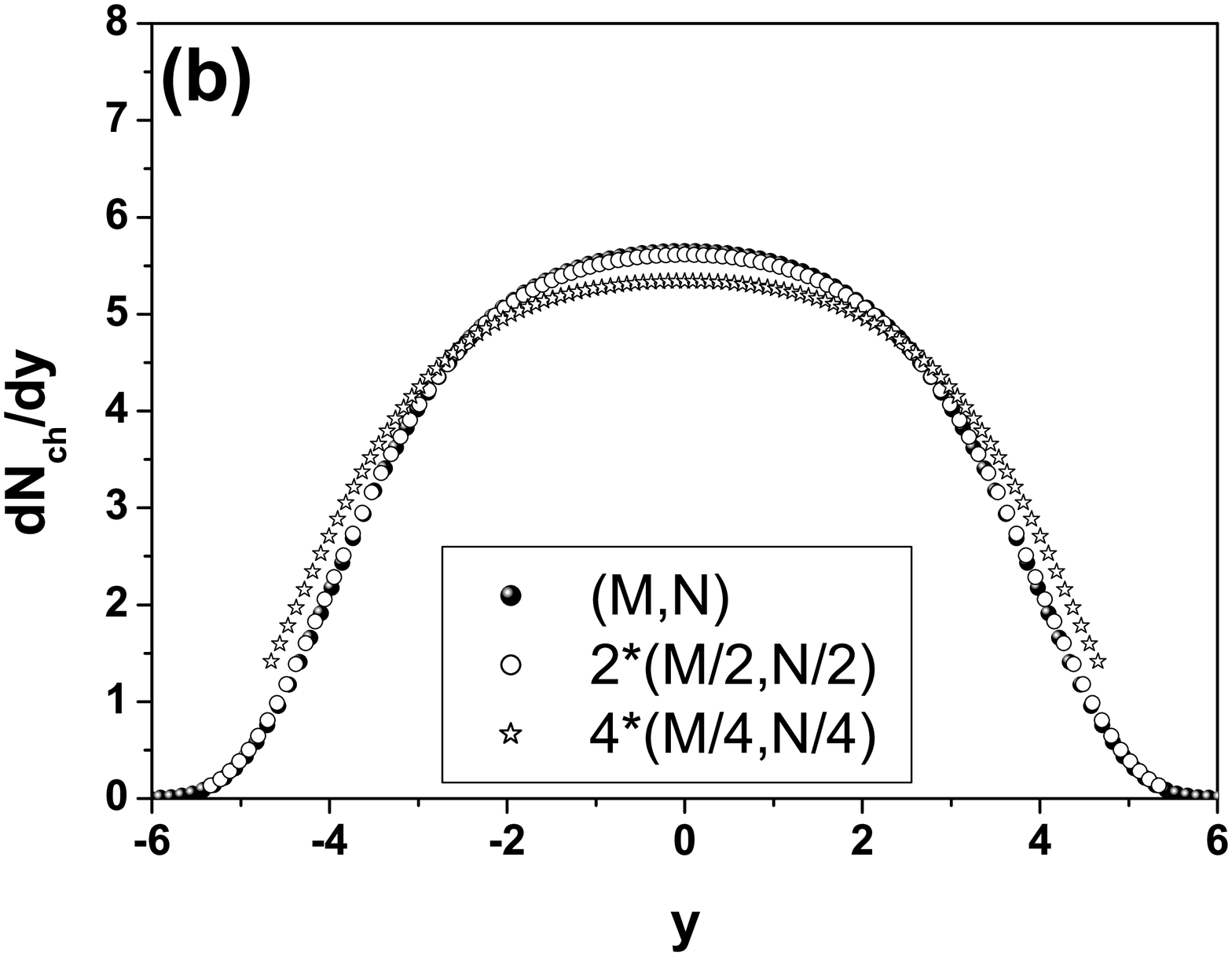, width=55mm}
     }
  \end{minipage}
\hfill
  \begin{minipage}[ht]{55mm}
    \centerline{
       \epsfig{file=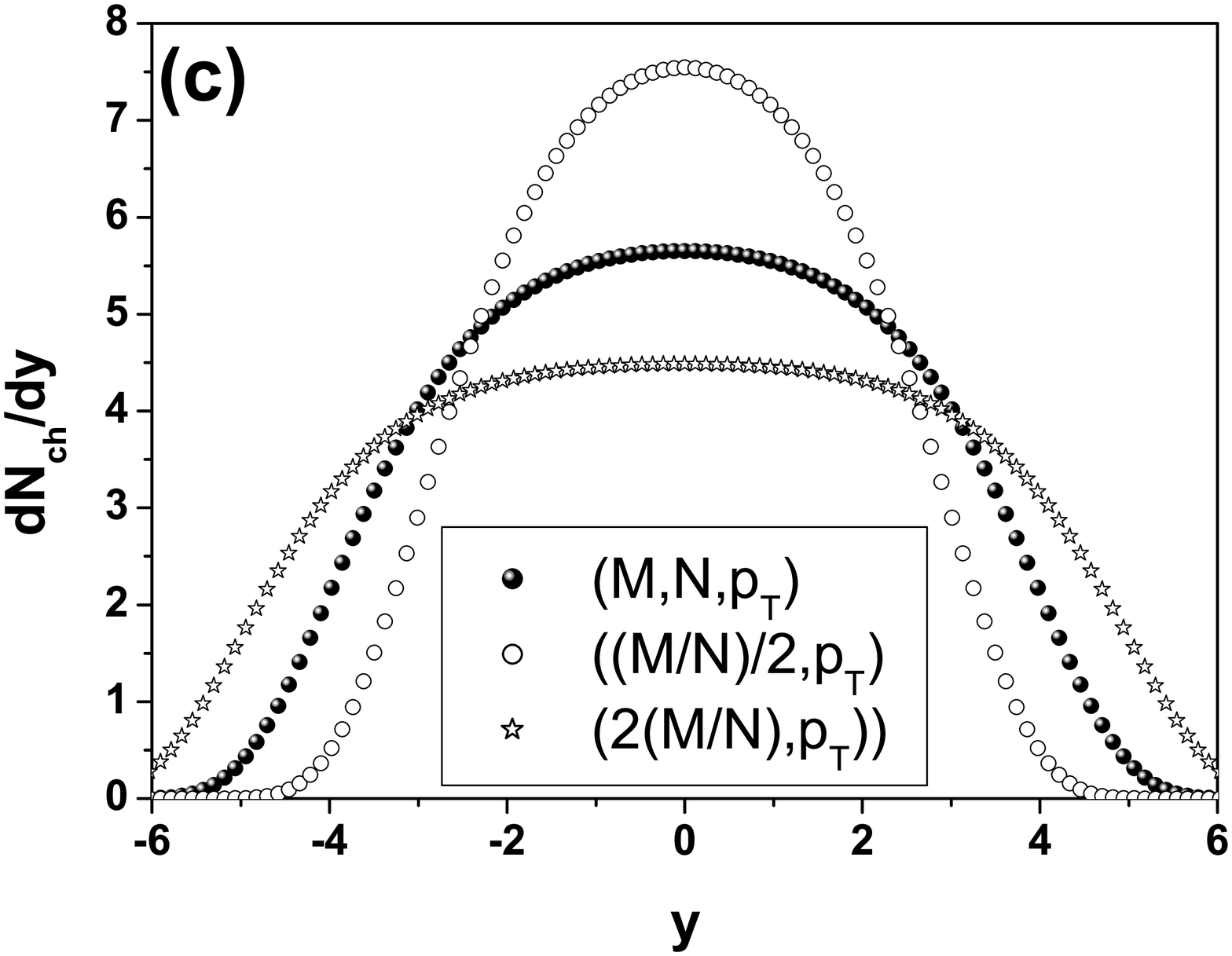, width=55mm}
     }
  \end{minipage}
\\
  \begin{minipage}[ht]{55mm}
    \centerline{
        \epsfig{file=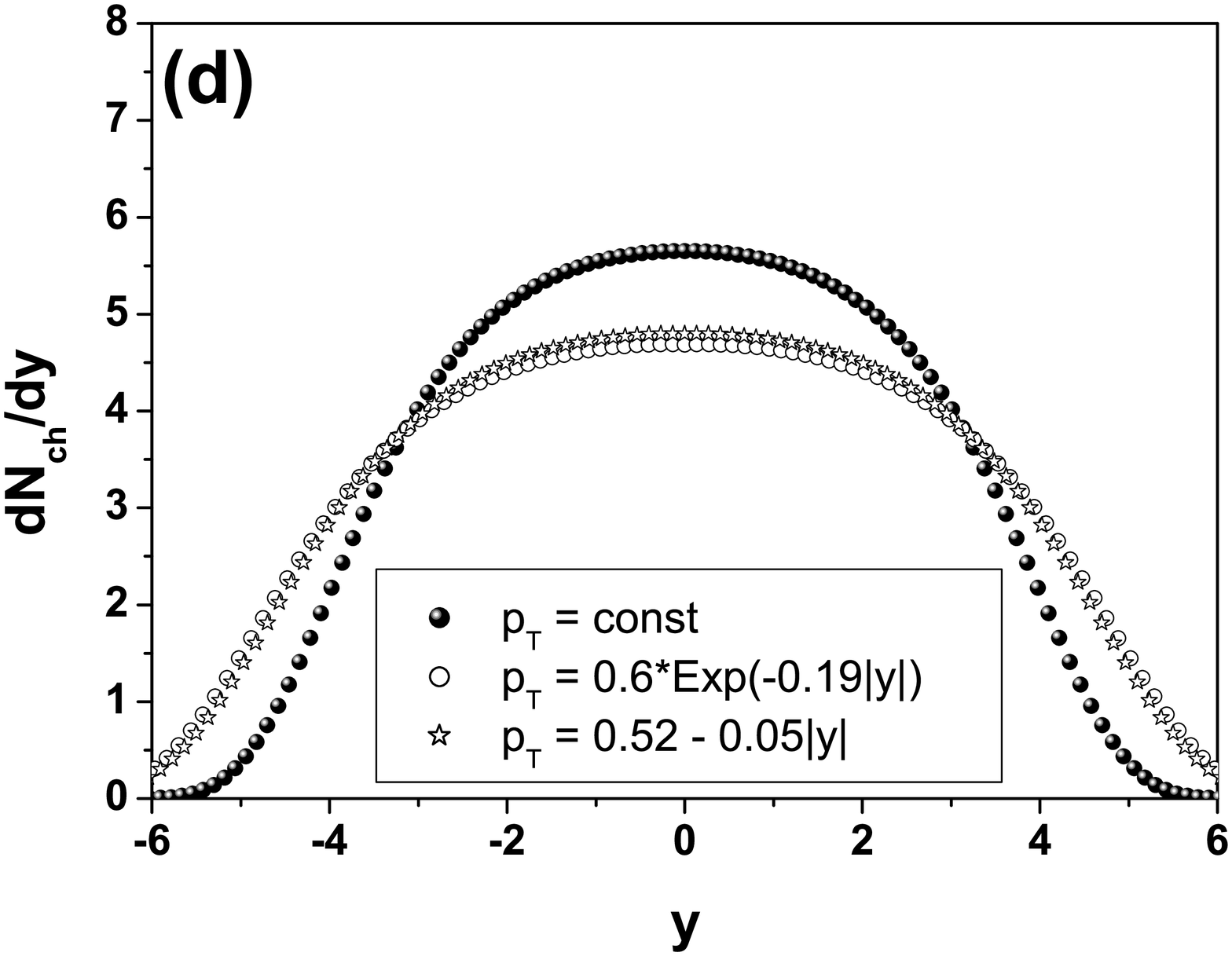, width=55mm}
     }
  \end{minipage}
\hfill
  \begin{minipage}[ht]{55mm}
    \centerline{
        \epsfig{file=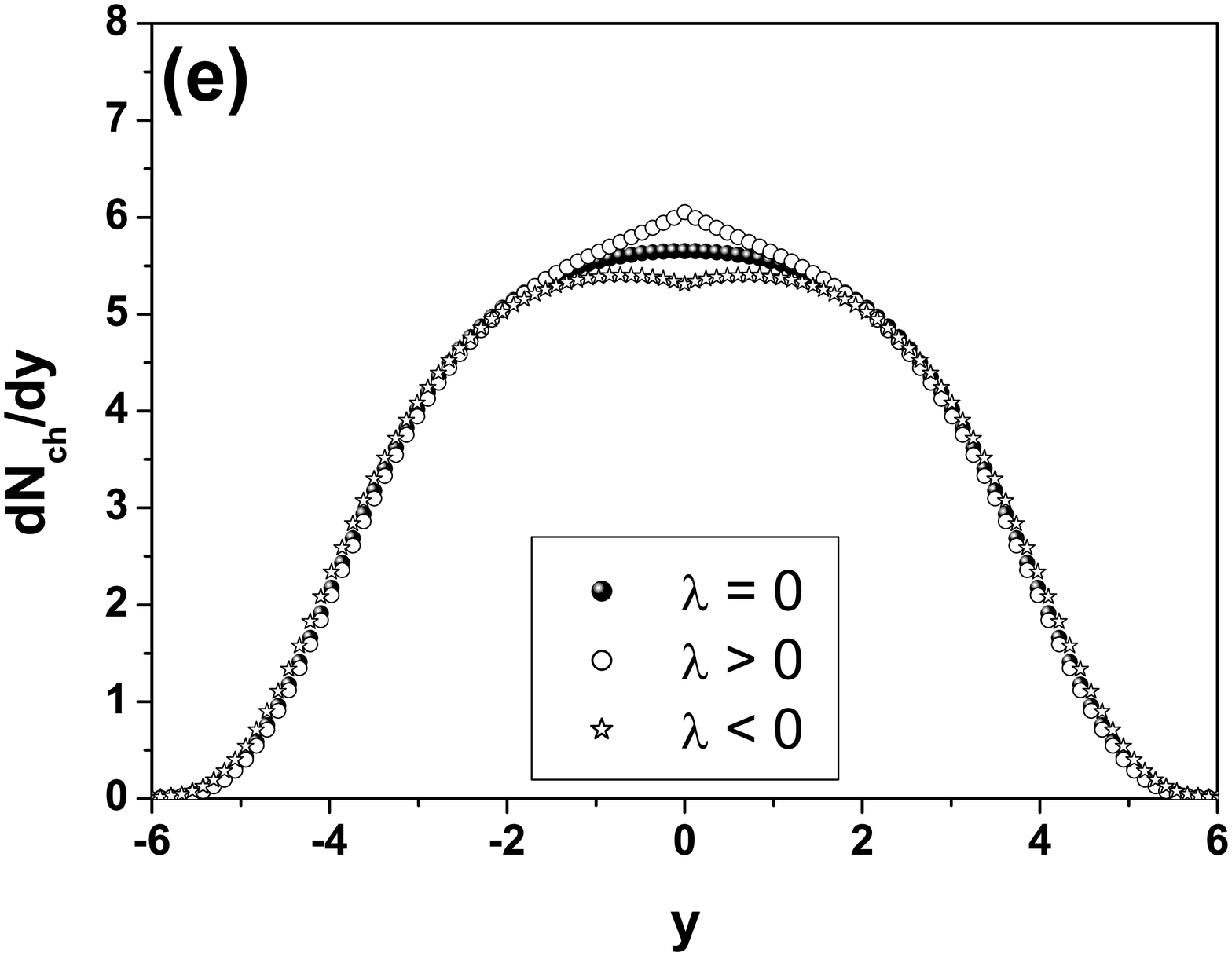, width=55mm}
     }
  \end{minipage}
\hfill
  \begin{minipage}[ht]{55mm}
    \centerline{
       \epsfig{file=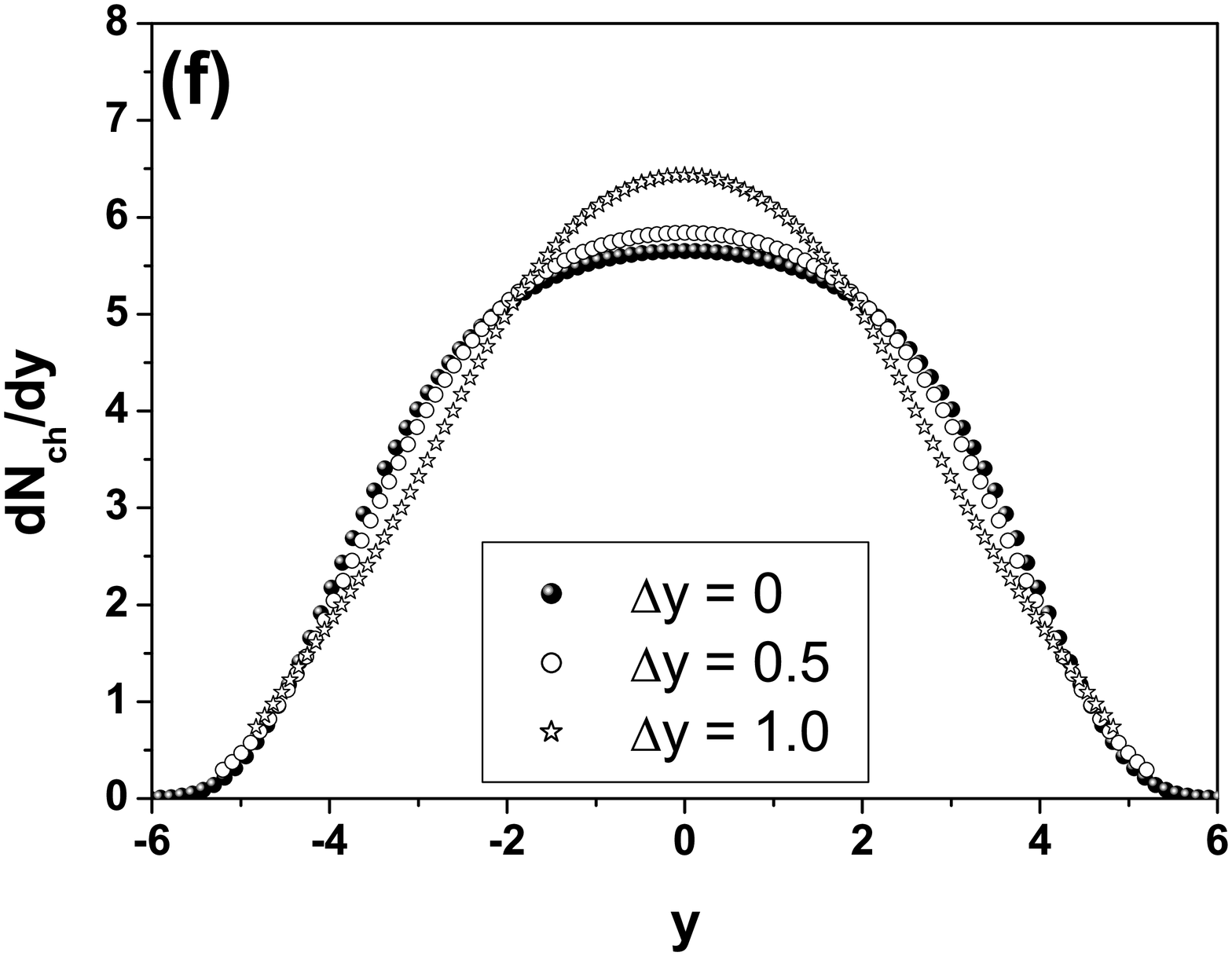, width=55mm}
     }
  \end{minipage}
  \caption{
\footnotesize Examples of properties of eqs. (\ref{eq:formulaq}) 
($W=200$ GeV, $N=60$, $\langle p_T\rangle = 0.4$ GeV/c). 
$(a)$ spectra obtained for different 
parameter $q$ practically coincide with extensive spectra produced by
masses $M^{\star}=M/(3-2q)$. For all remaining panels $q=1$.
$(b)$ quasiscaling for spectra obtained for
composition of smaller masses producing respectively smaller
number of secondaries with $M/N=$const. $(c)$
spectra with different values of $M/N$ ratio differ drastically.
$(d)$ the $p_T$ growing towards the center of rapidity phase
space has dramatic effect on spectra. $(e)$ the role of
momentum dependent residual interaction (cf. eq. (\ref{eq:lambda})) is
visualised. $(f)$ the effect of two sources separated 
in rapidity by $2\Delta y$ with combined energies and masses equal $M$ 
is visualised.} 
  \label{fig:Figure2}
\end{figure}

Finally, on Fig. \ref{fig:Figure1}c  we show fits to the recent NA49
data  \cite{DataAA} on $\pi^-$ production in $PbPb$ collisions at
three different energies (per  nucleon). The values of $(q;K_q=3\cdot
K^{(\pi^-)}_q)$ obtained are: $(1.2;0.33)$ for $17.3$ GeV,  
$(1.164;0.3)$ for $12.3$ GeV and $(1.04;0.22)$ for $8.6$ GeV. The
origin of $q>1$ in this case is not yet clear. The inelasticity seems
to grow with energy. It is also obvious that for higher energies
some new mechanism starts to operate because we cannot obtain in this
case agreement with data using only the energy conservation constraint.
The possibility offered in this matter by some closer inspection of
properties of eqs. (\ref{eq:formulaq}) and (\ref{eq:py}) are
displayed in Fig. \ref{fig:Figure2}. Whereas upper panels are
self-explanatory, the lower demand some attention. They show how
$p(y)$ can be distorted and it could be that one of the mechanisms
presented there is showing up in high energy NA49 data mentioned
above. Three mechanisms are listed there (only $q=1$ case is
considered). Fig. 2d shows the changes in $p(y)$
introduced by $y$-dependent $p_T$, especially when $\langle
p_T\rangle$ grows substantially towards $y=0$ (in such a way as to
keep the averaged $\langle p_T\rangle$ the same as original $0.4$
GeV/c). In this case one observes increase of $p(y)$ at small values
of $y$ but it seems to be too wide to explain the NA49 data.

Fig. \ref{fig:Figure2}e  shows result of the special kind of momentum
dependent residual interactions discussed in \cite{SG}. In our case
it would result in 
\begin{equation}
p(y) = \frac{1}{Z} \cdot \exp \left[ - \beta \cdot \mu_T \cosh y\,
-\, \lambda \cdot \mu_T \vert \sinh y \vert \right] \label{eq:lambda}
\end{equation}
with additional constraint imposed now on the modulus of the momenta:
\begin{equation}
\int_{-Y_m}^{Y_m}\, dy\, \mu_T \vert \sinh y\vert \, p(y)\, = \kappa
\cdot \frac{\sqrt{W^2 - M^2}}{N} . \label{eq:momentum} 
\end{equation}
Here $W = K\cdot \sqrt{s}$ and $M= \sum_{i}\sqrt{\mu_T^2 + p_{Li}^2}$ 
with $p_{Li} = \mu_T \sinh y_i$ (sum is over all produced
particles, $i=1,\dots,N$, and $\kappa$ is parameter telling us the
fraction of momentum turned into interaction). As it can be seen in Fig.
\ref{fig:Figure2}e, the effect of this interaction although visible is
very weak and probably negligible at present data. 

The most promising seems to be the conjecture that invariant energy
$W=K\cdot \sqrt{s}$ hadronizing into $N$ particles consists in
reality with two subsources of energies $W_1=W_2$, which are
displaced from $y=0$ by some rapidity $\Delta y$ (being a free
parameter here) and producing $N/2$ particles each,  $W_{1,2} = W/[2
\cosh (\Delta y)]$. As one can see, the obtained shape is very
similar to what is observed experimentally (cf. Fig. \ref{fig:Figure1}c).
 
\section{Summary}

In some types of data on multiparticle production processes one
frequently encounters ambiguity concerning the question, which of the
particular models used at that time is the correct one \cite{Chao}.
Such situation arises always when data contain only limited amount of
information. To select this information one has apply information
theory methods, which are widely known and used in other branches 
of science \cite{INFO}. The examples shown here show that information
theory ideas can be successfully used also to analyse data from
multiparticle production processes and that in this way one gets
highly model independent estimation of some quantities, in our
example it was inelasticity parameter $K$ \cite{qMaxEnt}. Actually,
in \cite{NC} we have attempted to fit single particle distribution
without {\it a priori} introducing neither inelasticity nor
fluctuations in mean multiplicity and found that it is possible {\it
only} with $q<1$. The reason turned out to be simple: in this case
the most important factor was decreasing of the available phase space
to mimic the action of the inelasticity and this can be done only
with $q<1$. The fit was not as good as shown here with notion of
inelasticity introduced explicitly but it was not very bad either.
As in \cite{WW,BeckC} we are stressing here the connection between
necessity to use nonextensive version of information theory and some
intrinsic fluctuations existing in the hadronizing system.
Finally, as was clearly displayed in Fig.\ref{fig:Figure2}, our
method seems to be also very useful in describing the  gross features
of the single particle spectra observed in heavy ion collisions. In
particular, it seems that with growing energy of collision there is
room  for some new mechanisms, not present at more elementary
nucleonic collisions. This point deserves therefore some special
scrutiny in the future. 

We would like to finish mentioning that there exists also example of
very successful use of the information theory approach to  describe
not only single but also double particle spectra, namely the so
called Bose-Einstein correlations observed between identical bosons
in all multiparticle production data, see \cite{OMT}. \\

\noindent
{\bf Acknowledgments:} GW is grateful to the organizers of the {\it
NEXT2003} for their support and hospitality. Partial support of the
Polish State Committee for Scientific Research (KBN) (grant 2P03B04123
(ZW) and grants 621/E-78/SPUB/CERN/P-03/DZ4/99 and 3P03B05724 (GW))
is acknowledged. 

%\newpage

\end{document}